\documentclass[twocolumn,amssymb,footinbib, nobibnotes, aps, pra, reprint, a4paper]{revtex4-1}

\usepackage{csquotes}

\usepackage{latexsym}
\usepackage{amssymb}
\usepackage{amsmath}
\usepackage{mathtools}
\usepackage{empheq}

\usepackage{physics}
\usepackage{siunitx}
\usepackage[british]{babel}
\usepackage{nicefrac}

\usepackage{graphicx, pgf}
\usepackage{svg}
\usepackage[hidelinks]{hyperref}
\usepackage{url}
\usepackage[labelfont=bf]{caption}

\usepackage{xcolor}

\usepackage{multirow}

\usepackage[%disable%
]{todonotes}

\usepackage[expansion=false, babel=true, protrusion = true]{microtype}

\newcommand\MPIDS{\affiliation{Max Planck Institute for Dynamics and Self-Organization, Göttingen, Germany}}
\newcommand\UGOE{\affiliation{Institute for Dynamics of Complex Systems, University of Göttingen, Göttingen, Germany}}

\newcommand{\mv}[1]{\mathbf{#1}}

\newcommand*{\kernhat}[2]{#2\kern#1\hat{\phantom{#2}}}

\usepackage{xspace}
\newcommand{\supprefmodeldefs}{Supp.~Sec.~1\xspace}
\newcommand{\suppreftolerance}{Supp.~Fig.~5\xspace}

\newcommand{\supprefmosaics}{Supp.~Fig.~4\xspace}

\usepackage[capitalise]{cleveref}

\newcounter{myfigpanel}[figure]
\newcounter{myfigpanelonly}[figure]

\newcommand{\panelletter}[1]{\refstepcounter{myfigpanel}\label{#1}\refstepcounter{myfigpanelonly}\label{onlyletter:#1}\Alph{myfigpanel}}
\newcommand{\panel}[1]{(\protect\panelletter{#1})}

\crefalias{myfigpanel}{figure}
\crefname{myfigpanelonly}{panel}{panels}

\makeatletter
\g@addto@macro\caption@prepareslc{%
    \renewcommand{\stepcounter}[1]{
        \caption@l@stepcounter{#1}
    }%
}
\makeatother

\begin{document}
\title{Sensitive particle shape dependence of growth-induced mesoscale nematic structure}
\author{Jonas Isensee}
\MPIDS\UGOE
\author{Philip Bittihn}
\email{philip.bittihn@ds.mpg.de}
\MPIDS\UGOE

\begin{abstract}
Directed growth, anisotropic cell shapes, and confinement drive self-organization in multicellular systems. We investigate the influence of particle shape on the distribution and dynamics of nematic microdomains in a minimal in-silico model of proliferating, sterically interacting particles, akin to colonies of rod-shaped bacteria. By introducing continuously tuneable tip variations around a common rod shape with spherical caps, we find that subtle changes significantly impact the emergent dynamics, leading to distinct patterns of microdomain formation and stability. Our analysis reveals separate effects of particle shape and aspect ratio, as well as a transition from exponential to scale-free size distributions, which we recapitulate using an effective master equation model. This allows us to relate differences in microdomain size distributions to different physical mechanisms of microdomain breakup. Our results thereby contribute to the characterization of the effective dynamics in growing aggregates at large and intermediate length scales and the microscopic properties that control it. This could be relevant both for biological self-organization and design strategies for future artificial systems.\end{abstract}

\maketitle
\section{Introduction}
Ensembles of passive particles can be grouped and characterized as classical phases of matter such as solids, fluids or glasses.
In active matter, the particles possess some form of inherent activity that drives the behavior.
Depending on the form and strength of activity, the resulting dynamic processes may cause aggregates to resemble classical phases~\cite{liverpoolViscoelasticitySolutionsMotile2001, hatwalneRheologyActiveParticleSuspensions2004, berthierNonequilibriumGlassTransitions2013, janssenActiveGlasses2019}
or to take on new forms such as self-propelled~\cite{vicsek_novel_1995} or rotating aggregates~\cite{tanOddDynamicsLiving2022}.
One type of activity that is not only highly relevant in biology but can also crucially influence the properties of soft matter is growth combined with cell division~\cite{ranftFluidization2010,gelimson2015_collective,hallatschekProliferatingActiveMatter2023}.
Energy is injected in the form of microscopic building materials and the required free energy to permit assembly,
leading to self-crowding induced confinement that can mediate self-organisation~\cite{araujoSteeringSelforganisationConfinement2023}.
Example systems are cells replicating within tissues, bacteria or, in future experiments, synthetic dividing droplets \cite{chengSyntheticBiologyEmerging2012, guindaniSyntheticCellsSimple2022, dreherLightTriggeredCargoLoading2021}.
While volume growth alone may be isotropic, (binary) division requires a symmetry breaking and therefore some anisotropy on the particle scale~\cite{hupeMinimalModelSmoothly2024}.
In fact, many microorganisms in nature such as bacteria expend energy to maintain their preferred shapes against external influences.
Actively maintained shapes combined with growth and division of the constituent cells is known to cause striking emergent behaviour:
Previous studies have explored the alignment dynamics of rod-shaped particles in simulations as well as in experiments with, e.g., E. coli bacteria~\cite{you_geometry_2018, boyer_buckling_2011, volfsonBiomechanicalOrderingDense2008, dellarcipreteGrowingBacterialColony2018}.
More recently, experimental studies have also linked cell shapes to competitive advantages~\cite{bergEmergentCollectiveAlignment2024a, changHowWhyCells2014}.
Many studies have focused on spherocylinder shapes resembling rod-shaped bacteria and consisting of a rectangular body and half-circle caps at opposing ends~\cite{boyer_buckling_2011,you_geometry_2018,Isensee2022,langeslay2024stress}.
It has been shown that such growing rods will align with the expansion flow when placed into micro-fluidic channels both in simulation and experiment~\cite{boyer_buckling_2011, bittihn_genetically_2020, Isensee2022}.
With sufficiently large rod aspect ratios, this alignment may even produce columnar tip-tip patterns toward the channel outlet.
The transition to perfect order is accompanied by a peculiar change in observed stress mechanics~\cite{Isensee2022}, indicating the stability of highly aligned patches even under excess compressive stress in the direction of cell orientation.
While the mechanics behind this phenomenon are not fully understood, it connects to the emergent dynamics in freely expanding colonies, which has been shown to spontaneously generate highly aligned microdomains when the rod aspect ratio is in the range that generates the stress inversion~\cite{you_geometry_2018, Isensee2022}.
In the context of channel geometry, the stability of the ordered state was already discussed in Ref.~\cite{boyer_buckling_2011} in the context of a \emph{buckling instability} which the authors attributed to the interactions of circular tips.

Despite the importance of cell shape in general and that of tip shape in particular, their influence on mesoscale structures such as microdomains has not been systematically explored.
Such information could be useful not only to understand principles for biological self-organization, but also determine necessary prerequisites for the design of artificial materials.
In this work, we explore this direction and investigate the role of tip shape.
We define parameterized shapes to explore small variations around the regular rod shape with circular tips and find a strong influence on the size of emergent structures.
Using a first-principles mathematical model, we illustrate how the distribution of observed microdomain sizes can be mapped to effective stability properties that hint at distinct breakup mechanisms for microdomains.

\begin{figure}[ht!]
    \centering
    \includegraphics[width=\linewidth]{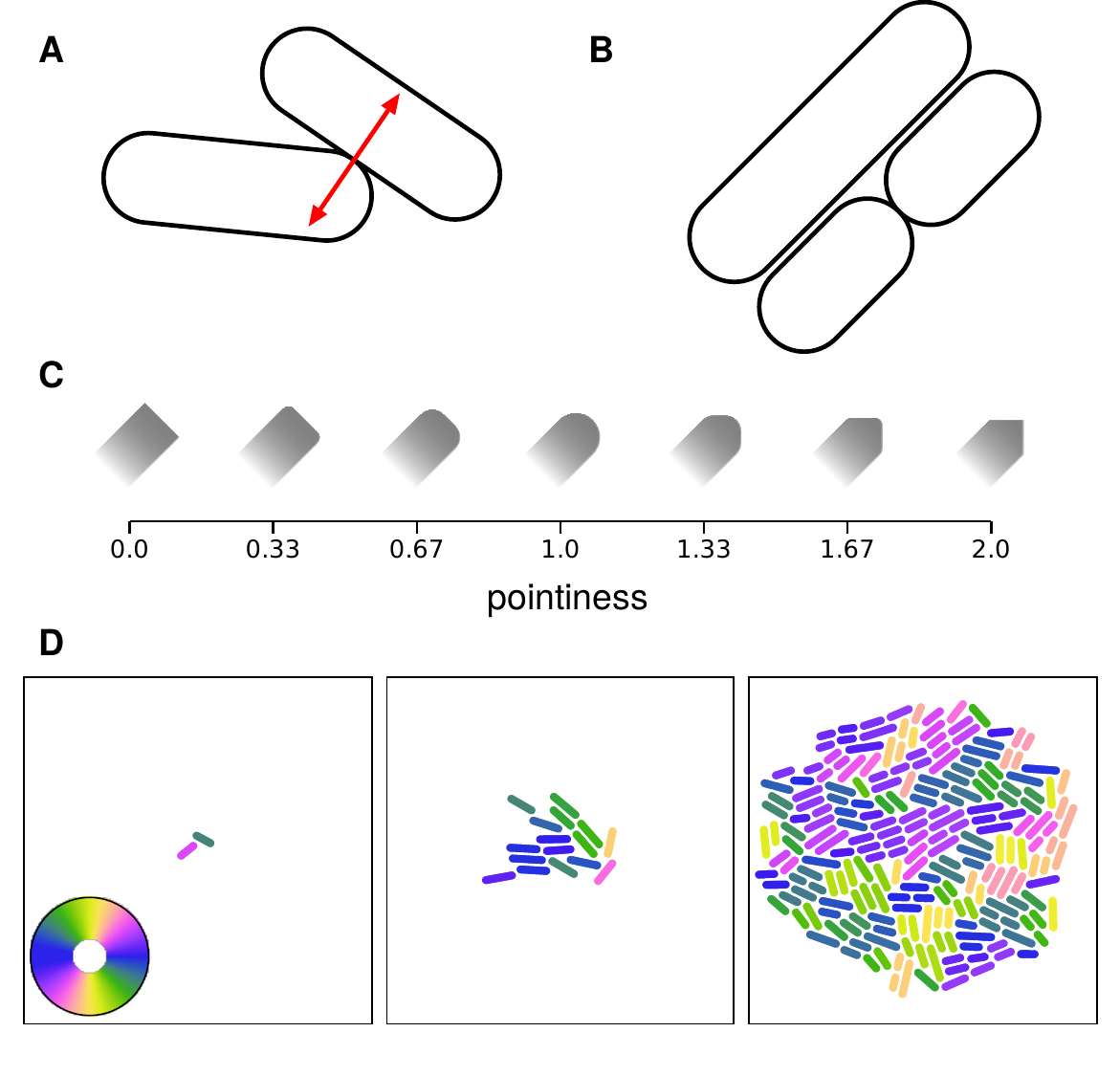}
    \caption{
    \panel{pan:stericrep} Illustrations of normal repulsion force between two touching agents and 
    \panel{pan:division} an agent dividing with inherited width, orientation and half the original length.
    \panel{pan:pointinessscale} Control parameter \emph{pointiness} to define modified tip-shape.
    \panel{pan:demosim} Snapshots at different times of the early stages of a simulation with coloring to emphasize (nematic) orientations as defined in the color wheel.
    }
\end{figure}
\section{Methods}

\subsection{Agent-based modeling}
\label{seg:abm}
For our numerical simulations, we consider agents of various shapes that grow in length and divide. Each agent occupies the space defined by its shape and the only interaction between agents is volume exclusion.
This is realized through the introduction of surface-normal repulsion forces whenever agents overlap as is illustrated in Fig.~\ref{pan:stericrep}.
The precise definition of the repulsion force depends on the shape details and is given in \supprefmodeldefs.

Simulations are set in the over-damped limit. 
This makes particle velocity $\mv{v}_i = \mu_i \mv{F}_i$ instantaneously defined by the sum of all pairwise interaction forces $\mv{F}_i$ and its mobility tensor $\mu_i$ as explained in more detail in \supprefmodeldefs. 

Our implementation of these models is available online as part of our in-house developed open-source framework for agent-based simulations of interacting particles InPartS~\cite{hupe_lukas_inparts_2022}.

\subsection{Agent shapes}
We define a set of particle models with the capacity for growth and division but varying shape details. 
The most common model for growing nematic bacteria is the rod model \cite{you_geometry_2018, dellarcipreteGrowingBacterialColony2018, Isensee2022, boyer_buckling_2011}.
There, particles consist of a fixed-width rectangular body and half-circle caps as evident in Fig.~\ref{pan:stericrep}.
A necessary requirement for all models with (indefinite) growth and binary divisions is that particle volume approximately doubles during its life-time. 
In the case of the rod model, the width and the caps stays fixed while the body length increases to reach its division aspect ratio $a_\text{d}$ as shown in Fig.~\ref{pan:division}.
Inspired by research on formation of smectics in liquid nematic crystals 
\cite{King2023-ss} we define a generalization of the rod model with a tip shape that can be continuously varied from flat to pointy as shown in Fig.~\ref{pan:pointinessscale}
with a control parameter \emph{pointiness} $\mathcal{P}$.
The pointiness takes values in the interval $\in [0,2]$ where a value of 1 recovers the original rod model.
For $\mathcal{P}\leq1$ the outline becomes a rounded rectangle with corner radius $r_c = \nicefrac{w}{2}\,\mathcal{P}$ equal to half the width $w$ times the pointiness.
For $\mathcal{P}\geq1$, the agents gain a (rounded) triangular cap with corner radius $r_c = \nicefrac{w}{2}(2-\mathcal{P})$ and side lengths scaled to preserve the bounding box. Further details are presented in \supprefmodeldefs.

\subsection{Setup}
A key challenge in characterizing emergent dynamics in models describing growth processes is to correctly identify or separate the emergent bulk dynamics from boundary effects (e.g. confinement)~\cite{Isensee2022} and history dependence (initial conditions).

Similar to previous research \cite{you_geometry_2018, dellarcipreteGrowingBacterialColony2018} we work without explicit confinement and consider a freely %
\footnote{%
Motion is opposed by surface friction.
Any expansion flow needs to be continuously driven by an internal pressure buildup.} %
expanding aggregate which, independent of the precise initial condition, quickly takes on a circular disk shape.
To also ensure access to stationary dynamics, we implement an indiscriminate removal of particles reaching a specified radial distance from the domain center.
For length scale comparisons across parameter variations, we define the area of newly created agents (approximated by their bounding rectangle) as a unit area.
Unless specified otherwise, the disk domains have a diameter of $150$ allowing for $\approx~10^4$ agents at a time.
An example time evolution for the early stages is shown in Fig.~\ref{pan:demosim}. Snapshots of full-size simulations are displayed in \cref{pan:snapshots}.

\section{Results}
The simulated dynamics spontaneously produce patches of highly aligned rods depending on division aspect ratio and tip shape. 
Examples of different combinations are shown in \cref{pan:snapshots}. Short regular rods (\textsc{i}) form no structures but short rectangular (\textsc{ii}) agents do form structures similar to much longer regular rods (\textsc{iv}) with $\mathcal{P} = 1$.
Increased pointiness reduces the size of emergent structures at large division aspect (\textsc{iii}) while the combination of low pointiness and large aspect yields the largest microdomains (\textsc{v}).
\begin{figure}[!h]
    \centering
    \includegraphics[width=\linewidth]{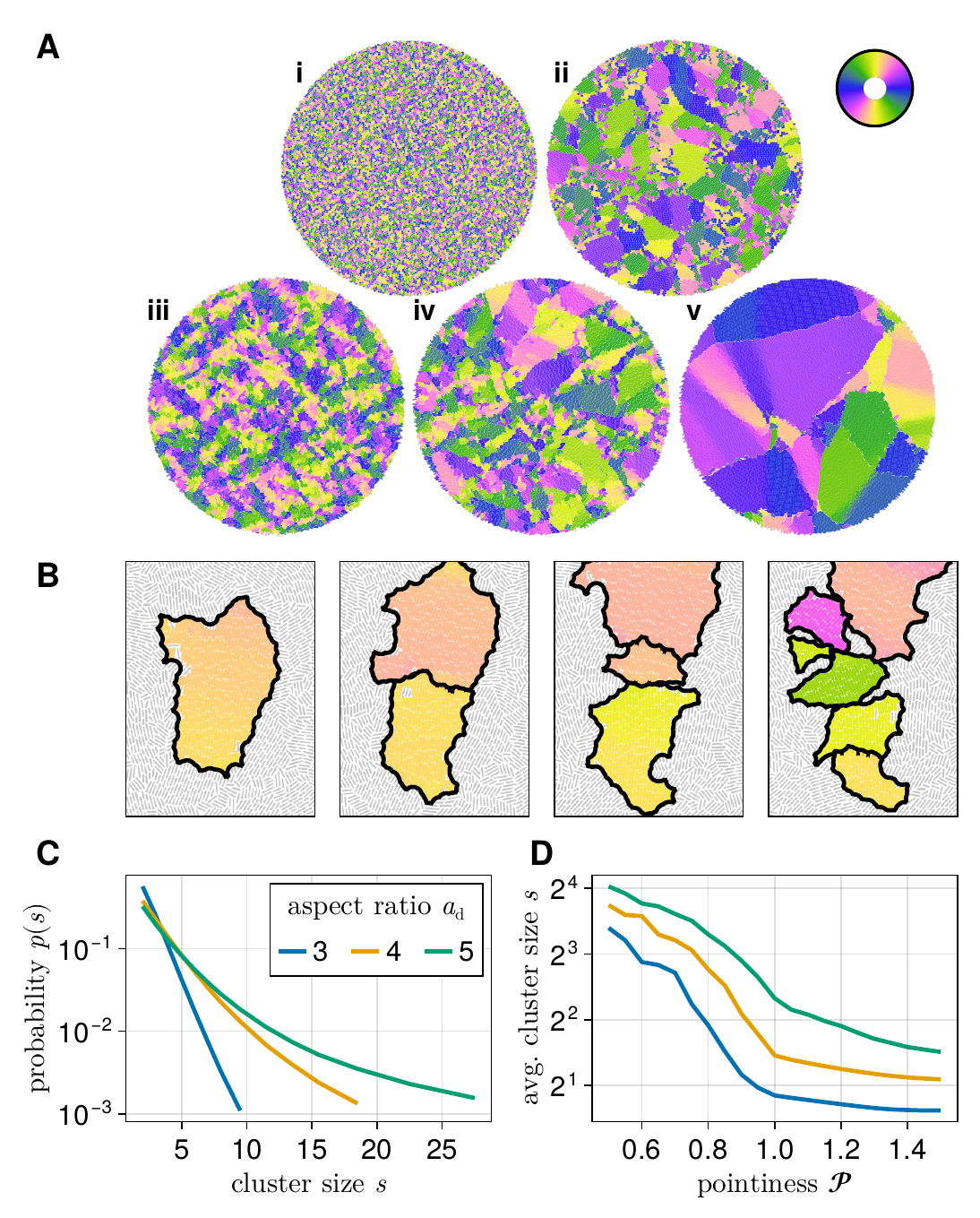}
    \caption{%
    \panel{pan:snapshots} Example snapshots for varied agent shapes.\
    The pointiness $\mathcal{P}$ has values $(1,\nicefrac{1}{2},\nicefrac{3}{2},1,\nicefrac{1}{2})$ in \textsc{i}-\textsc{v} and the aspect ratios $a_\text{d}$ are 2 in \textsc{i} \& \textsc{ii} and 5 in \textsc{iii}-\textsc{v}. The full time evolution of these examples is animated in the supplementary movies 1-5.
    \panel{pan:microdomainbreakup} A growing and breaking microdomain is tracked over time with $\Delta t=0.5$ between snapshots.
    \panel{pan:rodsizedemo} Cluster size distributions $p(s)$ for rods with aspect ratio $a_\text{d} =3,\, 4,$ and $5$. 
    \panel{pan:meansizepointiness} Average cluster size $s$ for varied pointiness $\mathcal{P}$ and aspect ratio $a_\text{d}$.
    }
\end{figure}

\subsection{Clustering}
As mentioned above and visually evident from the snapshots in \cref{pan:snapshots}, the agents group into highly aligned microdomains of characteristic parameter-dependent size.
To identify clusters, we construct an undirected graph with particles as vertices and add edges between interacting particle pairs whose orientations are approximately equal.
A standard connected-component analysis then yields the assignment into microdomains.
For regular rods a similar analysis was done by You et al.~\cite{you_geometry_2018}.
In general, the results of any clustering analysis will depend crucially on the choice of tolerance.
However, separately testing the effects of tolerance (done in \suppreftolerance) and subsequently keeping it constant for further analysis still allows for many conclusions to be made. All analyses shown here use a tolerance of $0.05\,\text{rad} \approx 1\deg$.
An example of a microdomain growing over time and subsequently breaking up into smaller pieces is shown in \cref{pan:microdomainbreakup}. While the clustering works on instantaneous configurations only, the continuity was recovered by tracking the constituent cells and their descendants in time.

Probability distributions of cluster sizes are shown in \cref{pan:rodsizedemo} for a range of parameters. 
For agents that form relatively small clusters (here $a_{\text{d}} = 3$) we find approximately exponentially distributed sizes. 
This is in line with the results by You et al.~\cite{you_geometry_2018}.
Interestingly, for large division aspect ratios $a_\text{d}$ the scaling gains heavy tails resembling a power law distribution
and as will be shown below, the same is true for low pointiness values.

To summarize the distributions, \cref{pan:meansizepointiness} shows the mean cluster size as a function of pointiness for division aspect ratios $a_\text{d}=3,\, 4$, and $5$.
The average cluster size increases the more boxy agents get ($\mathcal{P} \rightarrow 0$) leading to significant cluster formation even for the shortest ($a_\text{d} = 3$) agents. 
Pointy tip shapes inhibit the formation of large clusters.

Pursuing a better understanding of the mesoscopic dynamics, we will point out important aspects of the emergent size distributions.

\subsection{General properties}
In a first step, we view radially resolved stationary size distributions computed using the clustering algorithm introduced above.
Up to some boundary region, the size distribution is uniform in space as shown in Fig.~\ref{pan:radialdistr} where white indicates no observed clusters and all other bins are shaded according to a logarithmic scale of the observed number-density of clusters at the specified relative radial distance from the domain center.
Smallest clusters are the most frequent everywhere and the center of large clusters geometrically cannot reach the outer cutoff. 
Up to this geometric constraint, the distribution is not only homogeneous in space but also stationary without any meaningful transient.
This is visualized in Fig.~\ref{pan:timedistr} where a set of subsequent instantaneous size distributions are computed for a developing colony.
Initially, only very few agents are present but the domain becomes fully filled at time $t\approx 9$ and no meaningful changes of the distribution are observed after that.

Thirdly, when changing the overall system size, we find that the power-law type distribution tails, where present, scale accordingly. As a consequence, the average size of observed clusters remains approximately constant for division aspect ratio 3 and increases with domain size for higher division aspects as shown in Fig.~\ref{pan:avgsize_domain}.
The general shape of the distributions shown in Fig.~\ref{pan:distr_domain} on the other hand remains similar.
Only the \emph{cut-off} beyond the power-law-like tails moves to larger sizes.
This trend is difficult to see clearly from the log-log plot in the inset in \cref{pan:distr_domain}.
Hence, the main panel shows the distribution multiplied by the squared cluster size, approximately compensating for the power-law decay.

\begin{figure}
    \centering
    \includegraphics[width=\linewidth]{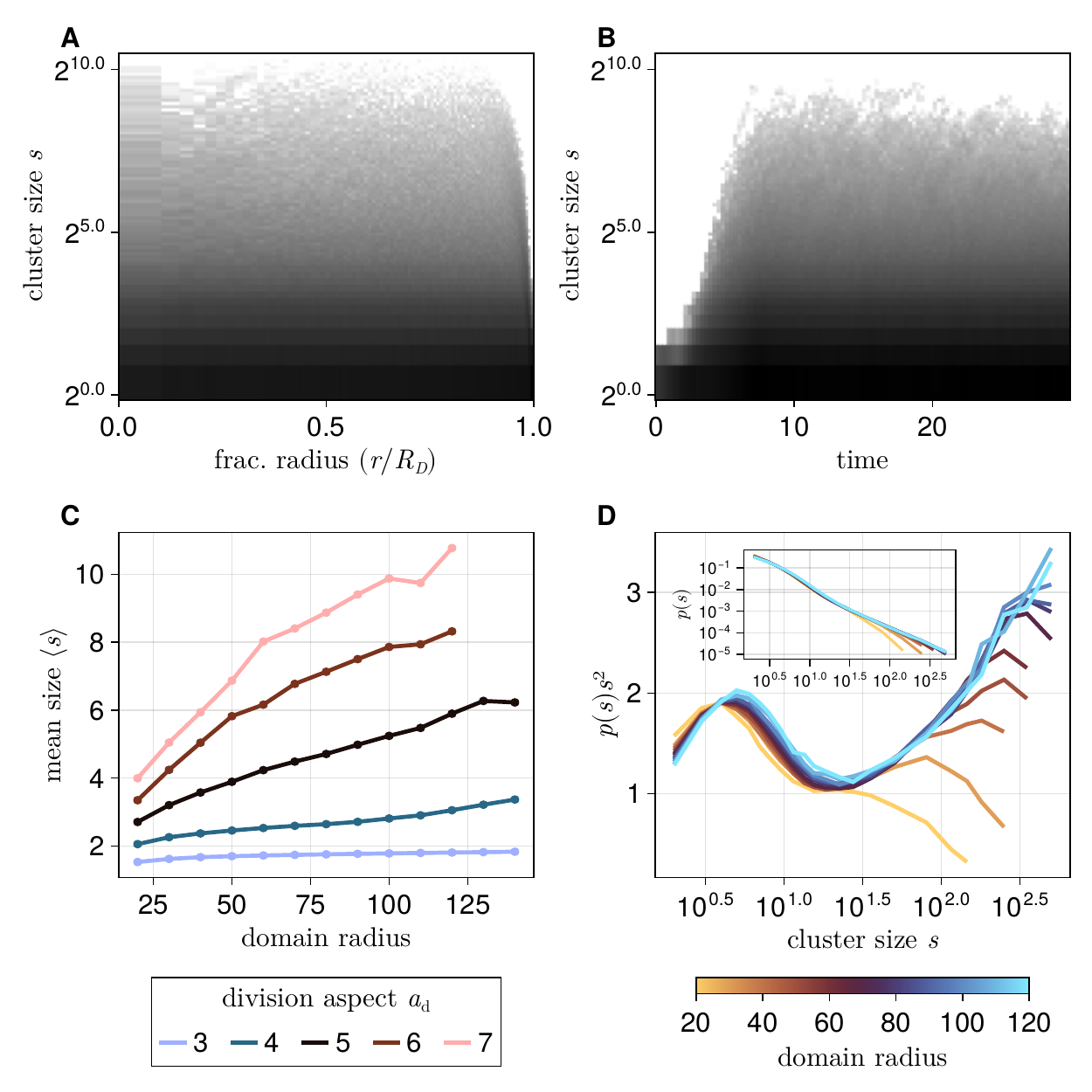}
    \caption{Size distribution properties of rodcells.
    \panel{pan:radialdistr} Cluster size distribution $p(s)$ as a function of radial position for division aspect $a_\text{d}=5$.
    \panel{pan:timedistr} Cluster size distribution over time with $a_\text{d}=5$.
    \panel{pan:avgsize_domain} Average cluster size for varied division aspect ratio $a_\text{d}$ and domain size.
    \panel{pan:distr_domain} Size distributions $p(s)$ for varied domain size at $a_\text{d}=6$, multiplied by $s^2$ to highlight tails.
    Regular distributions are shown in the inset.}
    \label{fig:moredistributions}
\end{figure}

\subsection{Shape variation}
With these general observations in hand, we can now turn to the parameter dependence of these distributions in more detail.
\cref{pan:rodlengthdistr} shows the measured sizes for rods with division aspects $a_\text{d}$ varied from 2 to 7.
To visually emphasize the differences between various distributions $p(s)$ we additionally display $p(s)s^2/Z$ in Fig.~\ref{pan:rodlengthdistr_ssq} with constant scaling $Z = p(s_{min})s_{min}^2$ and the quadratic size $s^2$ to bring the distribution tails into focus.
For all parameters the distributions are well approximated by the sum of an exponential decay and a power law
\begin{align}
    p(s) = c_1 e^{-s/\gamma_1} + c_2 s^{-\gamma_2}\label{eq:distributionfit}
\end{align}
except for the limits of the observed sizes where finite size and finite number effects dominate.

Short particles predominantly show exponential decay. The fit parameters in \cref{pan:rodlength_fitparams} for the power-law term yield large pre-factors but large decay exponents as well, indicating only small corrections at the smallest sizes.
An increase in aspect ratio $a_{\text{d}}$ first increases the decay length scale $\gamma_1$ but also drives a transition towards power-law scaling over multiple decades.
With increasing division aspect $a_{\text{d}}$ both the exponential decay length $\gamma_1$ and the power law exponent $\gamma_2$ move towards longer distribution tails.
Interestingly, this process saturates at division aspect $a_\text{d}\approx 5$ with the distributions seemingly converging both in length scale and amplitudes.
This is invariant under changes of the domain size in line with results from above.

\begin{figure}
    \centering
    \includegraphics[width=\linewidth]{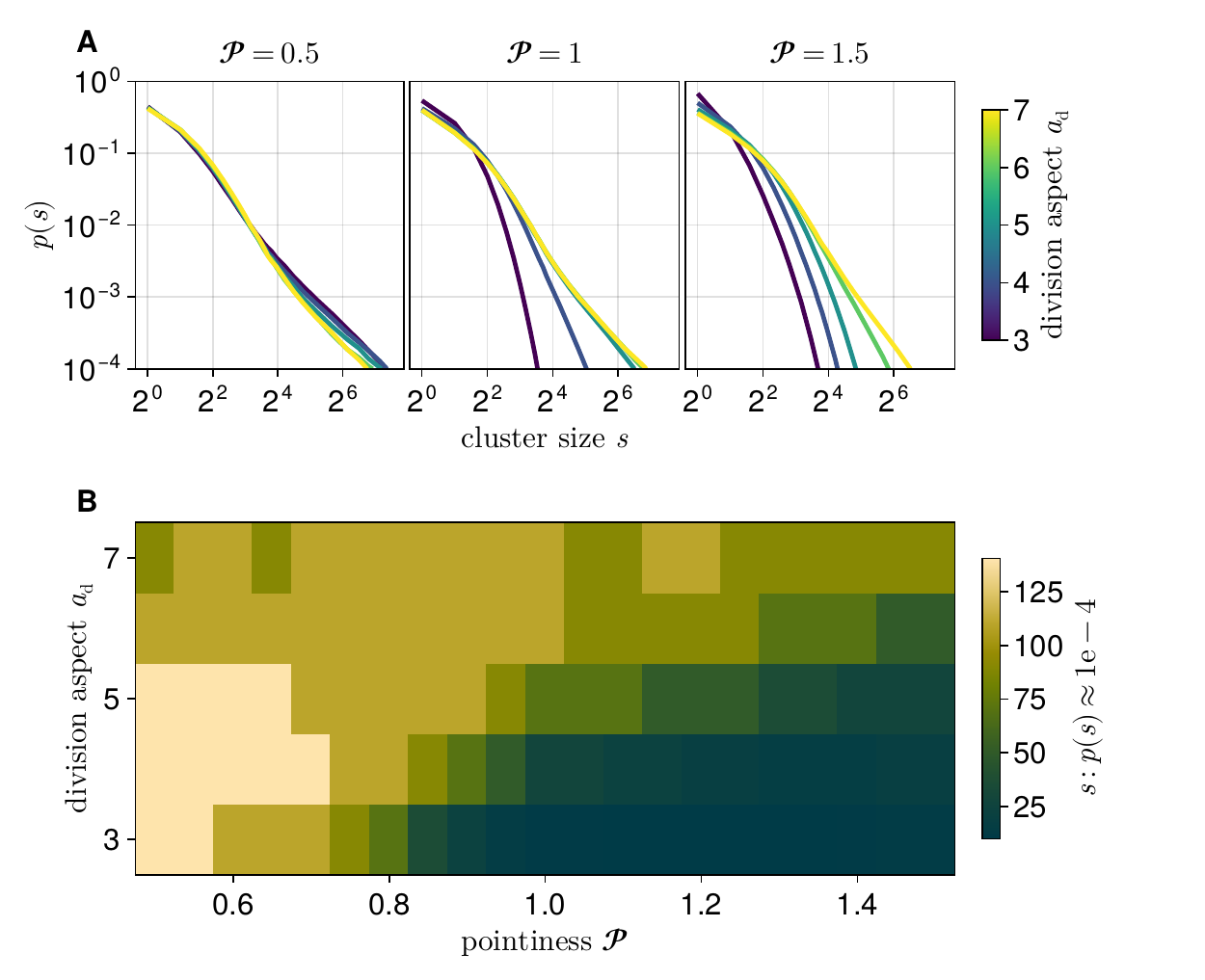}
    \caption{\label{fig:pointiness_distrs}
    Cluster size distributions $p(s)$ for varied pointiness $\mathcal{P}$.
    \panel{pan:pointiness_distr} Distributions multiplied for pointiness $\mathcal{P} \in [0.5, 1,1.5]$ split into separate panels. Coloring indicates division aspect ratio $a_{\text{d}}$.
    \panel{pan:pointiness_phasediagram} Distribution tail weight represented as the size $s$ at which the probability density crosses $p(s)=10^{-4}.$
    }
\end{figure}

Similar trends can be observed in \cref{fig:pointiness_distrs}
where both division aspect ratio $a_\text{d}$ and pointiness $\mathcal{P}$ are varied.
While pointy tip shapes $\mathcal{P}>1$ inhibit the formation of large clusters, an increased division aspect ratio still drives the distribution toward power-law tails.
For sufficiently low pointiness (e.g. $\mathcal{P}=0.5$) the distributions become heavy-tailed at all division aspect ratios.
The precise distribution shapes differ slightly from those of regular rods and are not described as well by \cref{eq:distributionfit}.
Instead, this trend is summarized in \cref{pan:pointiness_phasediagram} where tail weight is approximated using the cluster size $s$ for which $p(s) = 10^{-4}$ is reached. 
Parameters where exponential decay dominates appear dark, as the target probability is reached for comparatively small cluster sizes
and increased division aspect ratio or lowered pointiness drive toward larger sizes.
A non-obvious result is that the largest clusters are found in the bottom left corner of \cref{pan:pointiness_phasediagram} (for low pointiness and small division aspect ratio) rather than in the top left corner.
This indicates the existence of an optimal trade-off
between maximized aspect ratio and absolute width of the tips. These are not independent, as the average area is conserved.

\begin{figure}
    \centering
    \includegraphics[width=\linewidth]{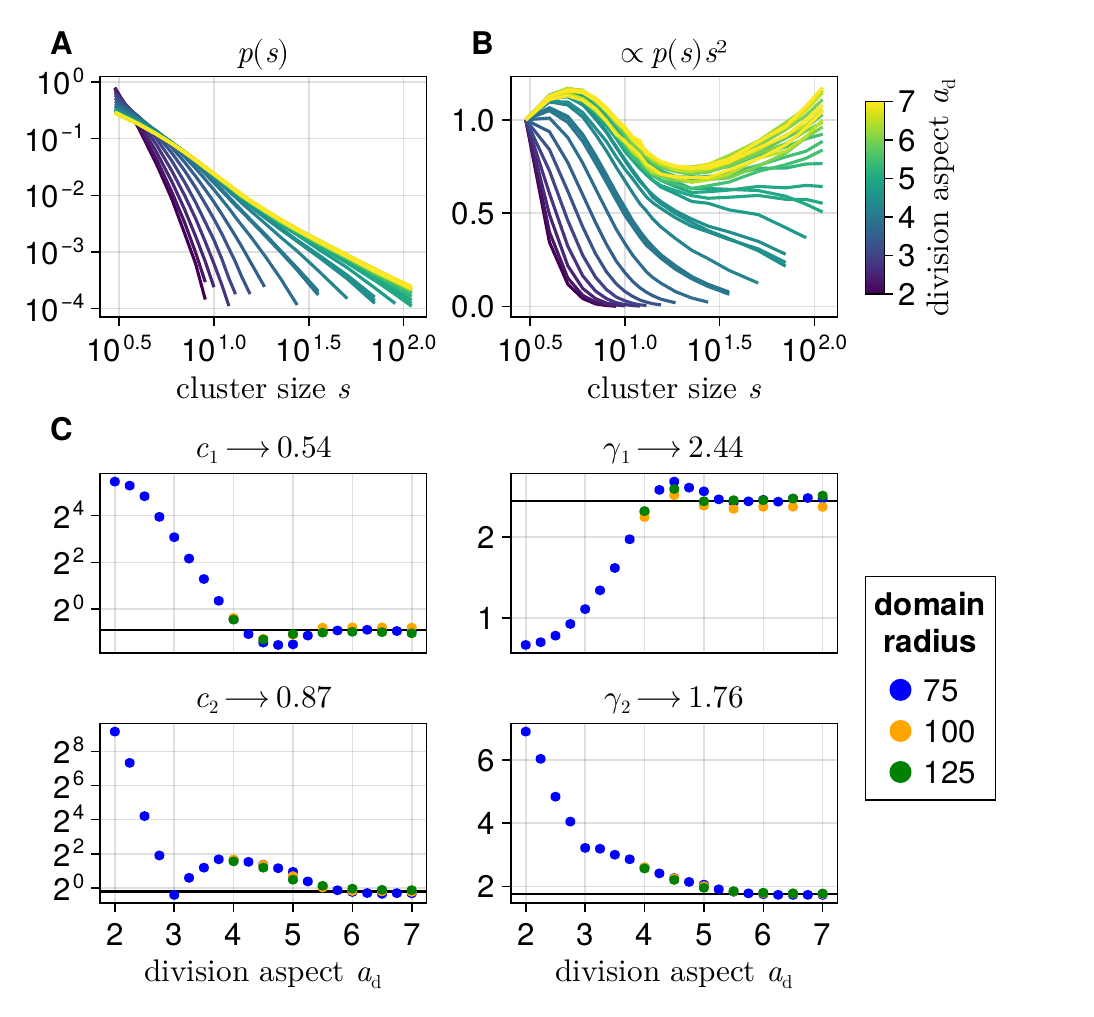}
    \caption{%
    \panel{pan:rodlengthdistr} Size distributions $p(s)$ for rods with varied division aspect ratio $a_{\text{d}}$.
    \panel{pan:rodlengthdistr_ssq} Distributions multiplied by $s^2$ and normalized to give 1 at the left-most data point to emphasize the distribution tails.
    \panel{pan:rodlength_fitparams} Fit parameters of Eq.~\ref{eq:distributionfit} to the data shown in \ref{pan:rodlengthdistr}.
    }
    \label{fig:rodlength_fits}
\end{figure}

\subsection{A master equation for microdomains}
\label{sec:master}
In the previous section we found that the emergent dynamics yields a stationary distribution of microdomain sizes.
Importantly, the shape of the distributions varies from exponential to predominantly power-law distributed.
To better understand the implications of this change, we set up a master equation for the likelyhood of finding microdomains of different sizes.
Microdomains consist of growing and dividing agents leading to exponential growth of each microdomain.
When considering the area fraction $A(s,t)\mathrm{d}s \propto s\, p(s,t)\mathrm{d}s$ of clusters of size $s$, time $t$, and number distribution $p(s,t)$,
growth can be expressed as an advection term with a velocity $s$ in size space due to
the exponential nature of growth.
Here, time is rescaled to absorb the growth-rate dependence.

The only additional process allowed for this calculation is microdomain breakup into equal halves with a size-dependent rate $\beta(s)$. In the area fraction description, this turns into a non-local coupling and the full equation yields
\begin{align}
    \partial_t A(s) + \partial_s (s A(s)) = - \beta(s)A(s) + 2\beta(2s)A(2s)\label{eq:distro_evo}
\end{align}
where the last term is evaluated at $2s$ and also has a factor 2 from the measure at $2s$.
Due to chosen units for $A(s,t)$ no additional factors are needed and the distribution stays normalized automatically.

Both the rate $\beta(s)$ and distribution $A(s,t)$ are a priori unknown.
However, we are specifically interested in understanding the dynamics involved in producing stationary exponential and power-law 
(number) distributions.
Assuming a specific stationary $A(s)$, we can solve for the corresponding $\beta(s)$.
For an exponential number density we require $A(s) \propto s e^{-s/a}$. Plugging this into \cref{eq:distro_evo} yields
\[ 0 = -2 + \frac{s}{a} - \beta(s) + 4\beta(2s)e^{-s/a}\]
and is solved by 
\begin{equation}
\beta(s) = \frac{e^{s/a}}{s^2} \left(1+ \sum_{n=1}^\infty \left(\frac{s^2}{2^n}- \frac{s^3}{8^na}\right)\exp\left(\frac{-s}{2^{n}a}\right)\right)\label{eq:beta_exp}.
\end{equation}
In constrast, requiring a number density with power-law decay $A(s) \propto s^{\gamma+1}$ leads to 
\[0 = -(\gamma+2) - \beta(s) + \beta(2s)2^{\gamma+2}\]
and is solved by the constant decay rate
\begin{equation}
\beta(s) \equiv \beta = \frac{\gamma+2}{2^{\gamma+2}-1}.\label{eq:beta_powerlaw}
\end{equation}

\begin{figure}
    \centering
    \includegraphics[width=\linewidth]{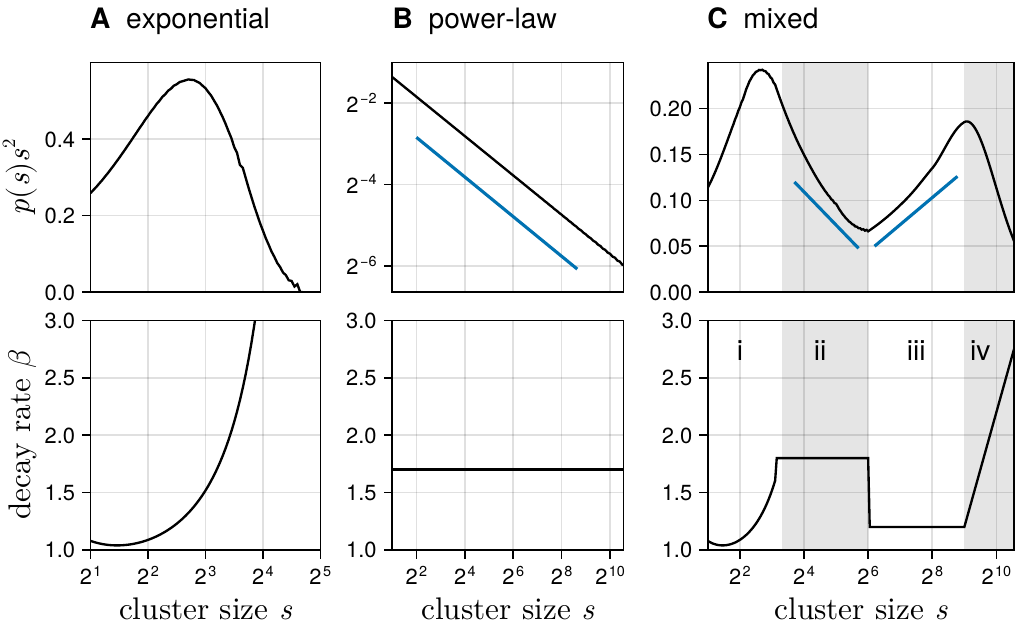}
    \caption{%
    Numerical steady state solutions of \cref{eq:distro_evo} for varied prescribed decay rate functions $\beta(s)$ in panels \panel{pan:exp}, \panel{pan:pow}, and \panel{pan:mixed}.
    Solutions, scaled with the squared size $s^2$, are shown above their corresponding $\beta$.
    Roman numerals are placed in \ref{pan:mixed} to label the (i) exponential, (ii) steep power-law, (iii) shallow power-law, (iv) long-range cut-off, regimes.
    Expected power-law scaling computed using \cref{eq:beta_powerlaw} is highlighted in blue.
    }
    \label{fig:expdistroanddecay}
\end{figure}

In the former case of an exponential number distribution we find a size-dependent decay rate. It is visualized in \cref{pan:exp} and features an increasingly large decay rate beyond the characteristic size $a$.
This is in line with the notion that microdomains beyond their stable size eventually buckle under their self-generated load.

Power-law distributions, shown in \cref{pan:pow}, on the other hand yield a constant decay rate independent of cluster size. This indicates that the buckling instability is no longer the dominant process and, instead, the dynamics is better described by breakup through stochastic input from the surrounding bath (of other microdomains).

An interesting observation in the data gathered from agent-based simulations is the bimodal nature of $p(s)s^2$ in e.g. \cref{pan:distr_domain}. By definition, this means that the local decay exponent varies above and below $-2$.
In the context of the decay model, this can be understood as an effective size-dependent variation in stability.
To emphasize this point, \cref{pan:mixed} shows a hand-crafted decay rate function $\beta(s)$. It consists of an exponential region (i, compare \cref{pan:exp}) for small sizes, two power-law regimes (ii \& iii) with varied expected exponent and a large size cut-off in (iv).
The analytically expected power-law scaling computed using \cref{eq:beta_powerlaw} were added in blue in both \cref{pan:pow,pan:mixed}.

At small cluster sizes, there is a natural limit imposed by the agent-based simulations that no clusters with fractional cell numbers may exist. The largest possible clusters are bounded by the finite simulation domain, however in practice no such clusters are ever observed.
This imposes a bound on the valid regime of power-law decay exponents.
The parameter fits in \cref{fig:rodlength_fits} revealed an asymptotic exponent $\gamma_2 < 2$.
If this scaling continued to infinite sizes, the average size would diverge and numerical observations should never reveal more than a single globally aligned patch.
As this is clearly not the case, an exponent $\gamma_2 < 2$ must be accompanied by a large size cut-off also added in \cref{pan:mixed}.

\section{Conclusion}

Our work shows that subtle differences in similarly spirited systems can yield surprising differences in the emergent dynamics.
Modifying the tip shape to be flat-spotted or more strongly pointed has a strong impact on the stability of tip-tip interactions that regularly occur after cell divisions and are a crucial ingredient in microdomain formation.
This extends the observations of Boyer et al. \cite{boyer_buckling_2011} who described the \emph{buckling instability} of ordered bacterial colonies due to circular tip interactions.

Lowering the pointiness and hence turning the rods into a rounded rectangle shape allows the agents to form long identically aligned columns prior to buckling.
This phenomenon enables the formation of large microdomains even with division aspect ratios that normally do not exhibit meaningful self-organization.

Increased pointiness on the other hand inhibits the formation of large clusters.
However, it enables an additional efficient packing configuration where neighboring rows of aligned agents are displaced by half a particle width.
This kind of behaviour has also been reported by King \& Kamien \cite{King2023-ss} and anecdotal evidence of this in our simulations is shown in \supprefmosaics.

In previous work~\cite{you_geometry_2018}, You et al. discovered that the characteristic length scale of microdomain size distributions depends on the ratio of an emergent bending stiffness of the ordered phase and an active stress extracted from an anisotropic average co-rotational stress tensor.
This is different in language but similar in spirit to the stability of ordered domains under anisotropic stress discussed in Ref.~\cite{Isensee2022}.
Notably, according to their analysis in Ref.~\cite{you_geometry_2018}, both emergent quantities are proportional to the excess density leaving the length scale constant. This is in agreement with our observation that the distribution of microdomain sizes is constant in the bulk and does not change with radial position. However, it is also incomplete as we find the average cluster size to be increasing with simulation domain size.

In this work, we found that the right combinations of tip-shape and aspect ratio can effectively inhibit self-buckling leading to a change in the size distributions towards power-law as argued in \cref{sec:master}. 
Beyond this point, only the total domain size appears to change the emergent distributions by setting a cut-off to the large-size power-law scaling.
This is a non-trivial observation.
Despite all interactions in the model being local pairwise repulsion forces, the domain boundaries manage to control a spatially homogeneous microdomain size distribution.
This rules out the simplest hypothesis that the total pressure field might be the sole mediator
and more analyses possibly on varied geometries could be done to determine the mediating field.

Due to our simple modeling assumptions, the phenomenology observed here can only scratch the surface of the dynamics possible in experimental analogues.
A peculiarity are the distinct disclination lines between microdomains.
That might change when rods were either allowed to bend with a finite stiffeness
or daughter cells optionally remained attached to one another after cell division~\cite{yaman_emergence_2019} where one could potentially observe a transition towards filamentous growing active nematics.

\section*{Data availability}
Simulation codes used for this study were built on top of the open-source framework InPartS.jl~\cite{hupe_lukas_inparts_2022}. The implementation of the agent-based models and solver for the master equation model will be made available alongside the final publication.

\section*{Acknowledgements}
We would like to thank our colleague, Lukas Hupe, for his valuable contributions to development and maintenance of our software infrastructure. His work has been an important part of the foundation on which this project relies.

\end{document}